\documentstyle[aps,preprint]{revtex}

\begin{document}

\title{Can Rotational Properties of General Relativistic Compact Objects Be
Predicted from Static Ones?}
\author{K.S Cheng\footnote{E-mail:hrspksc@hkucc.hku.hk}
    and T. Harko\footnote{E-mail: tcharko@hkusua.hku.hk}}
\address{Department of Physics, The University of Hong Kong,\\
Pokfulam Road, Hong Kong, P. R. China.}

\maketitle

\begin{abstract}
Abstract. We propose an approximate description of basic parameters (radius,
mass and oblateness) of general relativistic compact rotating objects in
terms of the parameters of the static configuration and of the angular
velocity only. The representation in terms of static properties is derived
using the condition of stationary equilibrium together with some
phenomenological assumptions. The predicted radius and mass of rotating
neutron star (described by some realistic equations of state) and strange star
(described by the bag model equation of state) are compared with data
obtained by numerical integration of gravitational field equation.The obtained
formulae also allow a simple derivation of the "empirical" equation relating
the maximum rotation frequency $\Omega _{max}$ of uniformly rotating star models to the mass and
radius of the maximum allowable mass configuration of the non-rotating model.
\end{abstract}

Pacs Numbers: 


\section{INTRODUCTION}

Rotation is a basic physical property of astrophysical objects. Study of
rotational properties of stars can lead to some restrictions imposed on the
nuclear equation of state of dense matter at densities larger than nuclear
density.Oscillations of rapidly rotating stars can become unstable hence
producing detectable gravitational wave emissions. 

In the past decades there were numerous attempts to construct analytic
models for rotating perfect fluid bodies in general relativity. But
obtaining an exact interior solution for a rotating body proved to be a
formidable task.The first step in this direction was the fundamental result
of Kerr [1] who obtained the solution for the vacuum domain, outside the
rotating star. It took no less than three decades of investigations to
obtain the first, highly idealized model of a general relativistic, thin
rotating disk of dust, by Neugebauer and Meinel [2]. Various schemes have
been developed for obtaining stationary and axisymmetric perfect
fluid solutions of the gravitational equations like Petrov type D, local
rotational symmetry, fluid kinematics, non-trivial Killing tensor, vanishing
Simon tensor, electric magnetic Weyl curvature, lagrangian or static-
stationary symmetry, geodesic eigenrays etc. (for a recent review of
rotating perfect fluid models in general relativity see [3]).

If in the case of rapidly rotating stellar configurations there are
still  many unsolved problems, a remarkable progress has been made in the
study of slowly and rigidly rotating perfect fluid configurations. By
casting the metric in the form
\begin{equation}
ds^{2}=e^{\nu }dt^{2}-e^{\lambda }dr^{2}-r^{2}\left[ d\theta ^{2}+\sin
^{2}\theta \left( d\phi -\omega dt\right) ^{2}\right] +O\left( \Omega
^{2}\right),   
\end{equation}
Hartle [4] obtained a formalism that proved to be very useful in many
investigations of the rotational properties of stars. However, this model,
which considers only first order corrections in $\Omega$, can not be used
to compute models of rapidly rotating relativistic stars with sufficient
accuracy.

On the other hand there has been recently a considerable advance in the
numerical understanding of rotating stars. Several high precision numerical
codes are now avalaible and it has been shown that they agree with each
other to remarkable accuracy (see [5] for a review of recent developments in
numerical study of rotatation, nonaxisymmetric oscillations and
instabilities of general relativistic stars).  

The non-sphericity of rapidly rotating stationary stellar configurations and
the complicated character of the interplay of the effects of rotation and of
those of general relativity seem not to permit a simple universal
description of rotating compact objects. However, Haensel and Zdunik [6] and
Friedman, Ipser and Parker [7] have found a simple relation connecting the
maximum rotation frequency $\Omega _{max }$ with the maximum mass
$M_{max }^{stat}$ and radius $R_{max }^{stat}$ of the static
configuration:
\begin{equation}\label{eq1}
\Omega _{max}=C_{S}\left( \frac{M_{max }^{stat}}{M_{\odot}}\right) ^{%
\frac{1}{2}}\left(\frac{R_{max }^{stat}}{10km}\right) ^{-\frac{3}{2}%
},
\end{equation} 
with $C_{S}$ a constant which does not depend on the equation of state of the
dense matter. The value of the constant $C_{S}$ has been obtained by fitting
equation (\ref{eq1}) with data obtained by numerically integrating the gravitational
field equations. It is given by $C_{S}=7200s^{-1}$ [7], $C_{S}=7700s^{-1}$ [8] or by 
$C_{S}=7730s^{-1}$ [9]. The empirical relation (\ref{eq1}) has been checked for many
realistic equations of state for neutron stars [8], [10]-[12] and for the case
of strange stars where the empirical formula also holds with a very good
precision, the relative deviations do not exceeding 2\% [9]. Equation
(\ref{eq1}), obtained on the basis of analyzing numerical solutions of the
gravitational field equations, provides an enormous simplification of the
problem of dynamical effects of rotation because the solutions of the
complicate general relativistic equations for a rotating star can be
replaced with the solutions of the much simpler TOV equation. Some attempts
to explain this empirical relation were not concluded with satisfactory
result. Weber and Glendenning [13], [14] used numerical models of slowly
rotating relativistic stars to show that the formula still hold,
but with $C_{S}=7500s^{-1}$.
Also in the slow rotation limit Glendenning and Weber [15]
derived a formula relating $\Omega_{max}$ to $M_{max}/R_{max}^{3}$, in terms of the
mass, equatorial radius, moment of inertia, angular momentum and quadrupole moment
of the maximally rotating configuration only. But it is not clear how the formula
(\ref{eq1}) follows from their results.Up to now a clear physical understanding
of this relation is still missing.

On the other hand a universal relation between the
maximum mass and radius of non-rotating neutron star configuration and the
mass $M_{max }^{rot}$ and radius $R_{max }^{rot}$ of the
configuration rotating with  $\Omega _{max}$ of the form
\begin{equation}\label{eq2}
M_{max }^{rot}=C_{M}M_{max }^{stat},R_{max }^{rot}=C_{R}R_{max}^{stat},
\end{equation} 
has also been found [12]. In (\ref{eq2}) $C_{M}$ and $C_{R}$
are specific equation of state dependent constants, whose values have
been calculated, for a broad set of realistic EOS, in [12]. Their mean values are
$C_{M}=1.18$ and $C_{R}=1.34$ [12].
The empirical constant $C_{S}$ can be obtained, within an approximation
better than $5\%$ , from the formula [12]
\begin{equation}
C_{S}\approx \left( \frac{C_{M}}{C_{R}^{3}}\right) ^{1/2}.
\end{equation}

The general validity of (\ref{eq1}) and (\ref{eq2}) suggests the possibility
that all basic physical
parameters of general relativistic rotating stellar objects (like mass and
radius) can be somehow related to the similar parameters of the static
configuration. It is the purpose of the present paper to propose a general
description of basic physical parameters (mass, radius and oblateness) of
compact general relativistic objects in terms of the physical
properties of the static configuration and of the angular velocity only. To
obtain the representation in terms of static physical parameters we use only
the general relativistic conditions of equilibrium for static and rotating
stars and some phenomenological assumptions.The resulted approximate mass and
radius formulae are compared with data obtained from numerical integration of
gravitational field equations in case of neutron stars described by realistic
equations of state and strange stars described by the bag model equation of
state.

The present paper is organized as follows.The general formalism allowing to obtain
mass and radius formulae for rotating general relativistic stars is presented
in Section 2. In Section 3 we apply our results to the case of neutron stars
described by realistic equations of state. In Section 4 we consider the case
of strange stars. We discuss and conclude our results in Section 5.

\section{THE GENERAL FORMALISM}

For a static equilibrium stellar type configuration, with interior described
by the metric
\begin{equation}
ds^{2}=e^{2\nu (r)}dt^{2}-e^{\lambda (r)}dr^{2}-r^{2}\left(d\theta ^{2}+\sin ^{2}\theta d\phi ^{2}\right),
\end{equation}
the condition of the hydrostatic equilibrium, which follows from the Bianchi identities, can be
written as
\begin{equation}
2\nu +2\int \frac{dp}{\rho +p}=const.,
\end{equation}
with $\rho$ and $p$ the energy density and the pressure of the matter respectively ( in the
present paper we shall use units so that $c=G=1$).

We shall also make the assumption
that the matter EOS is a one parameter dependent function,
\begin{equation}
\rho =\rho (n),p=p(n),
\end{equation}
with $n$ the proper baryon density. For such an equation of state the heat
function defined by
\begin{equation}
H(n)=\int \frac{1}{\rho (n)+p(n)}\frac{dp(n)}{dn},
\end{equation}
is a regular function of $n$.It can always be written in the form [16]
\begin{equation}
H(n)=\ln f(n),
\end{equation}
where
\begin{equation}
f(n)=\frac{\rho (n)+p(n)}{n},
\end{equation}
is the enthalpy per baryon.

As applied at the center and at the surface of the star respectively, the
hydrostatic equilibrium condition yields
\begin{equation}\label{0}
\left( 2\nu +2\ln f(n)\right)\mid _{center}=\left( 2\nu +2\ln f(n)\right)\mid _{surface}.
\end{equation}

At the vacuum boundary of the static star the Schwarzschild exterior
solution gives the metric and we have [17]
\begin{equation}\label{01}
2\nu _{S}=\ln \left( 1-\frac{2M_{stat}%
}{R_{stat}}\right),
\end{equation}
with $M_{stat}$ and $R_{stat}$ the mass and
radius of the static stellar configuration. We denote by $n_{C}$ and $n_{S}$
the baryon density at the center of the star and at the surface,
respectively. Therefore from (\ref{0}) and (\ref{01}) we obtain the following general and exact expression for the
mass-radius ratio of the static star:
\begin{equation}\label{00}
\frac{M_{stat}}{R_{stat}}=\frac{1}{2}\left[ 1-C\left( n_{C}\right) F\left(
n_{C},n_{S}\right) \right], 
\end{equation}
where we have denoted
\begin{equation}
F\left( n_{C},n_{S}\right) =\frac{f^{2}\left( n_{C}\right)}{f^{2}\left( n_{S}\right)},
\end{equation}
and
\begin{equation}
C\left( n_{C}\right) =\exp \left( \nu_{C}\right) =g_{00}\mid _{r=0}.
\end{equation}

For a given equation of state $C\left(n_{C}\right)$ is a function of the central density only.
From a physical point of view  $C\left(n_{C}\right)$ can be related via the
relation $\frac{1}{C}-1=z_{c}$ to the redshift $z_{c}$ of a photon emitted
from the center of the star.

For the mass of the star we can obtain another general representation by assuming
\begin{equation}\label{000}
M_{stat}=\frac{4\pi \rho _{spec}}{3}R_{stat}^{3}A\left( n_{C}\right),
\end{equation}
where $\rho _{spec}$ is a specific density that can be arbitrarily chosen
(for example it is the central density of the minimum mass configuration) and $%
A\left( n_{C}\right)$ is a function describing the effects of the variation
of the central density on the basic parameters of the stellar configuration.

The two unknown functions $C\left(n_{C}\right)$ and $A\left( n_{C}\right)$ can be determined by
fitting equations (\ref{00}) and (\ref{000}) with the exact values of the mass and radius
obtained by numerically integrating the TOV equation for a given equation of
state. Their knowledge allows us to construct exact mass and radius formulae
for sequences of static general relativistic stars having different central
densities. From equations (\ref{00}) and (\ref{000}) we obtain the radius and the mass of
the star in the form
\begin{equation}\label{*}
R_{stat}=\left( \frac{3}{16\pi }\frac{1}{A\left( n_{C}\right) \rho _{spec}}%
\right) ^{1/2}\left[ 1-C\left( n_{C}\right) F\left( n_{C},n_{S}\right) %
\right] ^{1/2},
\end{equation}
\begin{equation}\label{**}
M_{stat}=\frac{4\pi }{3}\rho _{spec}A\left( n_{C}\right) \left( \frac{3}{%
16\pi }\frac{1}{A\left( n_{C}\right) \rho _{spec}}\right) ^{3/2}\left[
1-C\left( n_{C}\right) F\left( n_{C},n_{S}\right) \right] ^{3/2} .
\end{equation}

Solving the equations $\frac{dR_{stat}}{dn_{C}}=0$ and $\frac{dM_{stat}}{%
dn_{C}}=0$ for the value of the central density $n_{C}$ will give, with the
use of (\ref{*}) and (\ref{**}) the maximum values of the radius and mass of the static star,
respectively.

To describe the interior of the rotating general relativistic star we shall
adopt the formalism presented in [16] and [9]. Under the hypothesis of
stationarity, axial symmetry and purely azimuthal motion a coordinate system 
$\left( t,r,\theta ,\phi \right)$ can be chosen so that inside the star the
line element takes the form
\begin{equation}
ds^{2}=N^{2}dt^{2}-B^{2}r^{2}\sin ^{2}\theta \left( d\phi -N^{\phi
}dt\right) ^{2}-A^{2}\left( dr^{2}+r^{2}d\theta ^{2}\right), 
\end{equation}
where $N$,$N^{\phi}$,$A$ and $B$ are functions of $r$ and $\theta$
only. As measured by the locally non-rotating observer the fluid 3-velocity
is given by $U=Br\sin \theta \frac{\Omega -N^{\phi }}{N}$ , where $\Omega =%
\frac{d\phi }{dt}$ is the angular velocity of a fluid element moving in the $%
\phi $-direction (physically it is the angular velocity as measured by an
observer at spatial infinity) [9],[16].

From the point of view of our phenomenological approach the most important
result is the equation of the stationary motion, which results from the
Bianchi identities and which for a rotating perfect fluid reduces to [16]
\begin{equation}\label{8}
\frac{1}{\rho +p}\frac{\partial p}{\partial x^{i}}+\frac{\partial \nu }{%
\partial x^{i}}-\frac{\partial }{\partial x^{i}}\left( \ln \Gamma \right) =-S%
\frac{\partial \Omega }{\partial x^{i}},
\end{equation}
where $S=\Gamma ^{2}\frac{A^{2}B}{N}Ur\sin \theta$ , $\nu =\ln N$ and $%
\Gamma =\left( 1-U^{2}\right) ^{-1/2}$. If $\Omega =const.$ (case called
uniform or rigid rotation) equation (\ref{8}) can be integrated to give the
following fundamental result describing the stationary equilibrium of a
rotating general relativistic star:
\begin{equation}\label{9}
\left( 2\nu +2\ln f(n)+\ln \left( 1-U^{2}\right) \right) \mid
_{center}=\left( 2\nu +2\ln f(n)+\ln \left( 1-U^{2}\right) \right) \mid
_{surface}.
\end{equation}

Equation (\ref{9}) is just the generalization to the case of rotation of the
well-known static Bianchi identity $2\nu +2\int \frac{dp}{\rho +p}=const.
$ [17] we have already used to describe static stellar configurations.

We assume that the vacuum boundary of the rotating star is described
by the Kerr metric [1] in the Boyer-Lindquist coordinates, which has the
form [18]:
\begin{eqnarray}
ds^{2} &=&\left( 1-\frac{2M_{rot}r}{r^{2}+a^{2}\cos ^{2}\theta }\right)
dt^{2}-\left( \frac{r^{2}+a^{2}\cos ^{2}\theta }{r^{2}+a^{2}-2GM_{rot}r}%
+r^{2}+a^{2}\cos ^{2}\theta \right) dr^{2}- \\
&&(r^{2}+a^{2}\sin ^{2}\theta )d\theta ^{2}-\left[ (r^{2}+a^{2})\sin
^{2}\theta +\frac{2M_{rot}ra^{2}\sin ^{4}\theta }{r^{2}+a^{2}\cos ^{2}\theta 
}\right] d\phi ^{2}-2\frac{2M_{rot}ra\sin ^{2}\theta }{r^{2}+a^{2}\cos
^{2}\theta }dtd\phi .  \nonumber
\end{eqnarray}

In this form the Kerr metric is manifestly axially symmetric and closely
resembles the Schwarzschild solution in its standard form.$M_{rot}$ is the
mass of the source and the parameter $a=-\frac{J}{M_{rot}}$ is the ratio
between the angular momentum $J$ and the mass$M_{rot}$ of the rotating star.

Let's apply equation (\ref{9}) at two points: at the center of the dense core and
at the pole of the rotating star. We denote by $C_{rot}^{(1)}$ the value of
the metric tensor component $g_{00}$ at the center of the star. At the polar
point $\theta =0$ and $r=R_{p}=const.$, where $R_{p}$ is the polar radius of
the star. We also have $U=0$ . Therefore at this point the line element is
given by
\begin{equation}\label{pol}
ds_{pol}^{2}=\left[ 1-\frac{2M_{rot}}{R_{p}}\left(1+a^{2}/R_{p}^{2}\right) ^{-1}\right] dt^{2}.
\end{equation}

Consequently from equations (\ref{9}) and (\ref{pol}) we obtain the following exact mass-polar radius relation for the
rotating general relativistic configuration:
\begin{equation}\label{11}
\frac{M_{rot}}{R_{p}}=\frac{1}{2}\left( 1+a^{2}/R_{p}^{2}\right) ^{-1}\left[
1-C_{rot}^{(1)}\left( n_{C},\Omega ,...\right) F\left( n_{C},n_{S}\right) %
\right]. 
\end{equation}

The function $F\left(n_{C},n_{S}\right)$ is identical to that of the
static case. Let's apply now equation (\ref{9}) for two points situated in the
equatorial plan of the rotating star: at the center of the star and at the
equator respectively. At the equator $r=R_{e}=const.$ and $\theta =\frac{\pi 
}{2}$ ($R_{e}$ is the equatorial radius of the star). For a uniform
rotation the rotation angle of the source/observer at the equator is $\phi
=\Omega t$ . Taking into account these results we obtain the Kerr metric at
the equator of the rotating star in the form:
\begin{equation}\label{13}
ds_{eq}^{2}=\left[ 1-\frac{2M_{rot}}{R_{e}}\left( 1+a\Omega \right)
^{2}-R_{e}^{2}\Omega ^{2}\left( 1+\frac{a^{2}}{R_{e}^{2}}\right) \right]
dt^{2}.
\end{equation}

With the use of (\ref{9}) and (\ref{13}) we obtain the following mass-equatorial radius
relation:
\begin{equation}\label{12}
\frac{M_{rot}}{R_{e}}=\frac{1}{2}\left[ \frac{1-C_{rot}^{(2)}\left(
n_{C},\Omega ,R_{e}\right) F\left( n_{C},n_{S}\right) -R_{e}^{2}\Omega
^{2}\left( 1+\frac{a^{2}}{R_{e}^{2}}\right) }{\left( 1+a\Omega \right) ^{2}}%
\right], 
\end{equation}
where a new function
\begin{equation}
C_{rot}^{(2)}\left( n_{C},\Omega ,R_{e}\right)=C_{rot}^{(1)}\left( n_{C},\Omega ,R_{e}\right)/ \left( 1-U^{2}\right) \mid_{r=R_{e}},
\end{equation}
has also been defined.

From Eqs. (\ref{11}) and (\ref{12}) we obtain the ratio $e$ of the polar and
equatorial radius (the oblateness) of the star in the form:
\begin{equation}\label{a}
e=\frac{R_{p}}{R_{e}}=\frac{1+\frac{a^{2}}{R_{p}^{2}}}{(1+a\Omega )^{2}}%
\frac{1-C_{rot}^{(2)}\left( n_{C},\Omega ,R_{e}\right) F\left(
n_{C},n_{S}\right) -R_{e}^{2}\Omega ^{2}\left( 1+\frac{a^{2}}{R_{e}^{2}}%
\right) }{1-C_{rot}^{(1)}\left( n_{C},\Omega ,R_{e}\right) F\left(
n_{C},n_{S}\right) }.
\end{equation}

In the case of the static star we have also proposed the alternative Eq.(\ref{000})
for providing another mass-radius relation. We shall generalize this
equation to the rotating case  by assuming that the following formula
relates the mass of the rotating star to its equatorial radius:
\begin{equation}\label{aa}
M_{rot}=\frac{4\pi \rho _{spec}}{3}R_{e}^{3}C_{rot}^{(3)}\left(
n_{C},\Omega ,R_{e}\right), 
\end{equation}
with $C_{rot}^{(3)}\left( n_{C},\Omega ,R_{e}\right)$ a function describing
the combined general relativistic effects of rotation and central density on
the mass of the star and with $\rho _{spec}$ the same specific density as
used in the static case. For $\Omega =0$, $R=R_{e}$ and Eq.(\ref{aa}) must
reduce to the static case equation (\ref{000}).

Equations (\ref{12}),(\ref{a}) and (\ref{aa}) give a complete and exact description of the mass and
radius of the rotating general relativistic star. But unfortunately the
present approach, which is basically thermodynamic in its essence, can not
predict the exact form and the values of the three unknown functions
entering in the formalism. The only thing we can do is to assume, also based
on the static case, some empirical forms for the functions $%
C_{rot}^{(i)}\left( n_{C},\Omega ,R_{e}\right) ,i=1,2,3$ and to check
whether the resulting formulae can give a satisfactory description of
rotating star configurations. Therefore in the following we shall use the
following five approximations:

i) As a first approximation we shall define the moment of inertia $I_{rot}$
of the rotating compact relativistic object via the Newtonian expression
\begin{equation}
I_{rot}=\frac{2}{5}M_{rot}R_{e}^{2}.
\end{equation}
In fact over a wide range of $M_{rot}$ and $R_{e}$ the corrections added by
general relativistic effects to the moment of inertia can be approximated by 
$I_{rot}=0.21\frac{M_{rot}R_{e}^{2}}{1-\frac{2M_{rot}}{R_{e}}}$ [15], but we shall
not use this representation. By adopting the Newtonian formula we obtain
\begin{equation}
a=-\frac{J}{M_{rot}}=-\frac{2}{5}R_{e}^{2}\Omega.
\end{equation}

ii) We assume that the function $C_{rot}^{(1)}\left( n_{C},\Omega
,R_{e}\right) $, describing the metric tensor component $e^{\nu }$ at the
center of the rotating star is given by 
\begin{equation}
C_{rot}^{(1)}\left( n_{C},\Omega ,R_{e}\right) =\left[ \left( 1-\frac{1}{2}%
g\left( \frac{M_{stat}}{R_{stat}}\right) R_{e}^{2}\Omega ^{2}\right) \right]
C\left( n_{C}\right) ,
\end{equation}
with $g\left( \frac{M_{stat}}{R_{stat}}\right) $ an EOS dependent function
given by 
\begin{equation}
g\left( \frac{M_{stat}}{R_{stat}}\right) =\frac{\alpha }{1-\frac{2M_{stat}}{%
R_{stat}}},
\end{equation}
and $\alpha $ a non-negative constant. $C\left( n_{C}\right) $ is also the
function corresponding to the static case.

iii) We assume that inside the rotating star $C_{rot}^{(2)}\left(
n_{C},\Omega ,R_{e}\right) $ is independent of the angular velocity $\Omega $ of the
rotating compact object and can be represented by the function corresponding to the
static case:
\begin{equation}
C_{rot}^{(2)}\left( n_{C},\Omega ,R_{e}\right) = C\left( n_{C}\right).
\end{equation}

iv) We suppose that
\begin{equation}
C_{rot}^{(3)}\left(
n_{C},\Omega ,R_{e}\right) =A\left( n_{C}\right)e.
\end{equation}
$A\left( n_{C}\right)$ is again the function corresponding to the static case.

With these four phenomenological assumptions Eqs. (\ref{12})-(\ref{aa}) lead to the following
representation of the basic physical parameters of the rotating general relativistic star:
\begin{equation}\label{b}
R_{e}=\left( \frac{M_{stat}}{R_{stat}}\right) ^{1/2}\left[ \frac{M_{stat}}{%
R_{stat}^{3}}-\frac{\alpha }{4}\Omega ^{2}\right] ^{-1/2},
\end{equation}
\begin{equation}\label{bb}
e=\frac{\left( 1+\frac{4R_{e}^{2}\Omega ^{2}}{25}\frac{1}{e^{2}}\right) }{%
\left( 1-\frac{2}{5}R_{e}^{2}\Omega ^{2}\right) ^{2}}\frac{\frac{2M_{stat}}{%
R_{stat}}-R_{e}^{2}\Omega ^{2}\left( 1+\frac{4}{25}R_{e}^{2}\Omega
^{2}\right) }{1-\left( 1-\frac{2M_{stat}}{R_{stat}}\right) \left( 1-\frac{1}{%
2}\frac{\alpha }{1-\frac{2M_{stat}}{R_{stat}}}R_{e}^{2}\Omega ^{2}\right) },
\end{equation}
\begin{equation}\label{bbb}
M_{rot}=\frac{4\pi \rho _{spec}}{3}A\left( n_{C}\right)
R_{e}^{3}e=M_{stat}\left( \frac{R_{e}}{R_{stat}}\right) ^{3}e.
\end{equation}

We shall also suppose that $\alpha$ is a universal
constants and we shall choose $\alpha\approx 5$.In this formulation of the general
relativistic problem of the rotation the oblateness parameter $e$ of the star is given
by the roots of the third order algebraic equation (\ref{bb}). 

The equatorial radius is defined only for values of the angular velocity
satisfying the condition $\frac{\alpha }{4}\Omega ^{2}\leq \frac{M_{stat}}{%
R_{stat}^{3}}$.Therefore for the maximum admissible constant angular
velocity of the maximally rotating star in uniform rotation we obtain the relation
\begin{equation}\label{xx}
\Omega \leq C\left( \frac{M_{max}^{stat}}{M_{\odot}}\right) ^{1/2}\left( 
\frac{R_{max}^{stat}}{10km}\right) ^{-3/2},
\end{equation}
where
\begin{equation}
C=\frac{2}{\sqrt{\alpha }}\sqrt{G}\frac{\left( M_{\odot}\right) ^{1/2}}{%
10^{9}}=10330.489s^{-1},
\end{equation}
for $\alpha =5$.

Equation (\ref{xx}) is very similar to
the ''empirical'' formula discussed in [1]-[7]. The coefficient of
proportionality in (\ref{xx}) is independent of the equation of state of the dense
matter, but its numerical value does not fit the calculated value.On the
other hand for $\Omega = C\left( \frac{M_{max}^{stat}}{M_{\odot}}\right)
^{1/2}\left( \frac{R_{max }^{stat}}{10km}\right) ^{-3/2}$ the radius of the
star tends to infinity.

An alternative expression can be obtained by
imposing the restriction
\begin{equation}
R_{e}^{max }\Omega _{max }\approx \beta \approx \sqrt{\frac{M_{max
}^{rot}}{R_{max }^{rot}}}\approx \sqrt{\frac{C_{M}M_{max }^{stat}}{%
C_{R}R_{max }^{stat}}}\approx \sqrt{\frac{M_{max }^{stat}}{M_{max }^{stat}%
}}, 
\end{equation}
where $\beta$ is the maxium allowable equatorial speed of the star
and we have also used the Newtonian force balance equation
between the gravitational and centrifugal force. Therefore we obtain
\begin{equation}\label{x}
\Omega _{max}\approx C\left( \frac{M_{max }^{stat}}{M_{\odot}}\right)
^{1/2}\left( \frac{R_{max}^{stat}}{10km}\right) ^{-3/2}\left[ 1+\frac{%
4M_{max }^{stat}}{\alpha \beta ^{2}R_{max}^{stat}}\right] ^{-1/2},
\end{equation}

It is interesting to note that the values of $2M_{max}^{stat}/R_{max}^{stat}$
are in a narrow range of (0.467, 0.667) [8], [12]. Taking for $2M_{max}^{stat}/R_{max}^{stat}$ a mean
value of 0.58 [12] it follows that $\beta \approx 0.54$ and we find  
\begin{equation}
\Omega _{max}\approx C_{S}\left( \frac{M_{max}^{stat}}{M_{\odot}}\right)
^{1/2}\left( \frac{R_{max}^{stat}}{10km}\right) ^{-3/2},
\end{equation}
with $C_{S}=7708s^{-1}$, value which coincides with that proposed in [8]
and differs only within $1\%$ from the value $C_{S}=7840s^{-1}$ obtained in [9].

By taking for twice of the ratio  of the maximum static mass and radius a mean value
of 0.58 [12] and considering $\beta \approx 0.54$, we obtain
\begin{equation}
\Omega _{max}\approx C_{S}\left( \frac{M_{max}^{stat}}{M_{\odot}}\right)
^{1/2}\left( \frac{R_{max}^{stat}}{10km}\right) ^{-3/2},
\end{equation}
with $C_{S}=7708s^{-1}$,value wich coincides with the value obtained in that differs within $5\%$ from the value
$C_{S}=7840s^{-1}$ obtained in [20].

The maximum radius of the maximally rotating configuration can be
obtained from $R_{max }^{rot}\approx \frac{\beta }{\Omega _{max }}$, and
is given by
\begin{equation}\label{ww}
R_{max }^{rot}\approx C_{R}\left( \frac{M_{max }^{stat}}{R_{max }^{stat}}%
\right) R_{max }^{stat},
\end{equation}
where
\begin{equation}
C_{R}\left( \frac{M_{\max }^{stat}}{R_{\max }^{stat}}\right) =\frac{\beta }{%
10km}\frac{1}{C_{S}}\sqrt{\frac{M_{\odot }}{10km}}\left( \frac{M_{\max }^{stat}%
}{R_{\max }^{stat}}\right) ^{-1/2},
\end{equation}

The mass $M_{maxrad}^{rot}$ of the maximal radius rotating neutron star
follows from (\ref{bbb}) and is given by
\begin{equation}
M_{maxrad}^{rot}=C_{RM}\left( \frac{M_{max }^{stat}}{R_{max }^{stat}}%
\right) M_{max }^{stat},
\end{equation}
where
\begin{equation}
C_{RM}\left( \frac{M_{max }^{stat}}{R_{max }^{stat}}\right)
=C_{R}^{3}\left( \frac{M_{max }^{stat}}{R_{max }^{stat}}\right) e_{max
}^{rad}\left( \frac{M_{max }^{stat}}{R_{max }^{stat}}\right),
\end{equation}
\begin{equation}
e_{max }^{rad}\left( \frac{M_{max }^{stat}}{R_{max }^{stat}}\right) =%
\frac{1}{\left( 1+\frac{2}{5}\beta ^{2}\right) ^{2}}\frac{1-\left( 1-\beta
^{2}\right) ^{\gamma }\left( 1-\frac{2M_{max }^{stat}}{R_{max }^{stat}}%
\right) -\beta ^{2}\left( 1+\frac{4}{25}\beta ^{2}\right) }{1-\left( 1-\frac{%
2M_{max }^{stat}}{R_{max }^{stat}}\right) \left[ 1-\frac{\alpha \beta ^{2}%
}{2}/\left( 1-\frac{2M_{max }^{stat}}{R_{max }^{stat}}\right) \right]}.
\end{equation} 

The maximum mass $M_{max }^{rot}$ of the rotating star can be obtained
from the equation
\begin{equation}
\frac{\partial M}{\partial \Omega }\mid _{\Omega =\Omega _{max}}=\frac{%
M_{stat}}{R_{stat}}\frac{\partial }{\partial \Omega }\left(
R_{e}^{3}e\right) \mid _{\Omega =\Omega _{max }}=\frac{M_{stat}}{R_{stat}}%
\left( 3R_{e}^{2}\frac{\partial R_{e}}{\partial \Omega }e+R_{e}^{3}\frac{%
\partial e}{\partial \Omega }\right) \mid _{\Omega =\Omega _{max }}=0,
\end{equation} 
leading to 
\begin{equation}\label{www}
e_{max }\left( \Omega _{max }\right) =\frac{C\left(
M_{stat},R_{stat}\right) }{R_{e}^{3}\left( \Omega _{max }\right) }=\frac{%
C_{M}\left( \frac{M_{stat}}{R_{stat}}\right)R_{stat}^{3}}{R_{e}^{3}\left( \Omega _{max }\right)},
\end{equation}
where $C_{M}\left( \frac{M_{stat}}{R_{stat}}\right)$ is
a dimensionless EOS dependent function. Therefore for the maximum mass of the maximally
rotating configuration we obtain 
\begin{equation}\label{w}
M_{max}^{rot}=C_{M}\left( \frac{M_{stat}}{R_{stat}}\right)M_{max}^{stat}.
\end{equation}

Equations (\ref{ww}) and (\ref{w}) show the existence of a proportionality relation
between maximum mass and radius of the rotating and non-rotating configuration,
respectively, as has already been suggested, on an empirical basis,in [12].

An investigation of 12 EOS performed in [12] for 12 realistic EOS of nuclear
matter led to a (mean) value of $C_{M}=1.1807$ , while the same calculation
performed by us with the use of data presented in [20] for other 14
different EOSs gives $C_{M}=1.177$ , leading to a mean value of $C_{M}=1.1790
$ for the 26 considered EOSs.

Eqs. (\ref{b})-(\ref{bbb}) lead, for $\beta \approx 0.54$ and
$2M_{max}^{stat}/R_{max}^{stat} \approx 0.58$ to values of $e_{max}=0.365$ and
$C_{M}=1.233$.

Therefore we may conclude that the proportionality
between the maximum mass of the rotating star and the maximum mass of the
static configuration is universal, being with a very good approximation independent of the equation of state
of dense matter. On the other hand the relation between the radius of the rotating
and non-rotating configuration is EOS dependent,the coefficient of proportionality
slightly decreasing with the increase of the mass-radius ratio of the static
star.

\section{APPLICATIONS TO NEUTRON STARS}

The correct mathematical and physical modelling of millisecond pulsars can be done
only in the framework of general relativistic equilibrium models for rapidly
rotating neutron stars.Such models are solutions of the Einstein's equations
for the axisymmetric stationary gravitational field and they must be constructed
numerically. Recently several independent numerical codes have been developed by
different groups of researchers and have been used to obtain rapidly rotating
neutron star models based on a variety of realistic equations of state.Hence
a large amount of numerical data is now available. 

In the present Section we shall apply the results of the phenomenological formalism
presented in the previous Section to the case of neutron stars described by
realistic equations of state. The data are selected from the paper by Cook, Shapiro
and Teukolsky [20],who constructed general relativistic rotating star sequences for
14 nuclear matter equations of state. Detailed data are presented only for 5 equations
of state. For the sake of comparison we have used the equations of state denoted
A [21], AU [22], FPS [23] and L [24]. At low densities all these equations of state
employ the Feynman,Metropolis and Teller [25] EOS and then join onto the Baym,Bethe and
Pethick [26] EOS up to neutron drip.The equations of state are given in a tabular
form and small changes in the way the tabulated equation of state is constructed
do have a small effect on the resulting neutron star model.

In Tables 1-4 we present the comparison of the basic physical parameters of
rotating stars obtained, for these four equations of state, with the help
of Eqs. (\ref{b})-(\ref{bbb}) and by 
numerically integrating the gravitational field equations.

\begin{table}[h]
\begin{center}
\begin{tabular}{|c|c|c|c|c|c|c|}
$\Omega (s^{-1})$ & $R_{num}(km)$ & $R_{calc}(km)$ & $\frac{R_{calc}}{R_{num}}$ & $%
M_{num}(M_{\odot }$ & $M_{calc}(M_{\odot})$ & $\frac{M_{calc}}{M_{num}}$ \\\hline 
0 & 9.586 & 9.586 & 1.0 & 1.40 & 1.40 & 1.0 \\\hline 
3244.1 & 9.763 & 9.899 & 1.014 & 1.4030 & 1.423 & 1.014 \\\hline
5018.9 & 10.06 & 10.393 & 1.033 & 1.4077 & 1.456 & 1.034 \\\hline 
6136.6 & 10.38 & 10.876 & 1.047 & 1.4123 & 1.483 & 1.050 \\\hline 
6940.0 & 10.74 & 11.340 & 1.055 & 1.4169 & 1.506 & 1.063 \\\hline 
7544.8 & 11.14 & 11.777 & 1.057 & 1.4214 & 1.525 & 1.073 \\\hline 
7953.7 & 11.56 & 12.125 & 1.048 & 1.4252 & 1.538 & 1.079 \\\hline 
8236.4 & 12.00 & 12.397 & 1.033 & 1.4285 & 1.547 & 1.083 \\\hline 
8431.1 & 12.49 & 12.602 & 1.008 & 1.4313 & 1.554 & 1.085 \\\hline 
8590.6 & 13.72 & 12.781 & 0.931 & 1.4340 & 1.559 & 1.087
\end{tabular}
\caption{Comparison of the radius $R_{calc}$ and mass $M_{calc}$
obtained from Eqs.(\ref{b})-(\ref{bbb}) with the
radius $R_{num}$ and mass $M_{num}$ obtained by numerically integrating the gravitational field
equations [20] for EOS A [21]: $1.4M_{\odot}$ normal mass sequence}
\end{center}
\end{table}

\begin{table}[h]
\begin{center}
\begin{tabular}{|c|c|c|c|c|c|c|}
$\Omega (s^{-1})$ & $R_{num}(km)$ & $R_{calc}(km)$ & $\frac{R_{calc}}{R_{num}}$ & $%
M_{num}(M_{\odot })$ & $M_{calc}(M_{\odot })$ & $\frac{M_{calc}}{M_{num}}$ \\\hline 
0 & 10.85 & 10.85 & 1.0 & 1.4 & 1.4 & 1.0 \\\hline 
2883.8 & 11.08 & 11.259 & 1.016 & 1.4031 & 1.424 & 1.015 \\\hline
4112.1 & 11.36 & 11.736 & 1.033 & 1.4065 & 1.449 & 1.030 \\\hline 
5033.6 & 11.71 & 12.268 & 1.047 & 1.4103 & 1.472 & 1.044 \\\hline 
5712.1 & 12.10 & 12.791 & 1.057 & 1.4140 & 1.491 & 1.055 \\\hline 
6174.6 & 12.49 & 13.233 & 1.059 & 1.4173 & 1.504 & 1.061 \\\hline 
6544.0 & 12.94 & 13.649 & 1.054 & 1.4206 & 1.515 & 1.066 \\\hline 
6841.8 & 13.51 & 14.035 & 1.039 & 1.4239 & 1.522 & 1.069 \\\hline 
7016.9 & 14.06 & 14.286 & 1.016 & 1.4263 & 1.527 & 1.070 \\\hline 
7165.0 & 15.45 & 14.519 & 0.939 & 1.4287 & 1.530 & 1.071
\end{tabular}
\caption{Comparison of the radius $R_{calc}$ and mass $M_{calc}$ obtained
from Eqs.(\ref{b})-(\ref{bbb}) with the
radius $R_{num}$ and mass $M_{num}$ obtained by numerically integrating the gravitational field
equations [20] for EOS FPS [22]: $1.4 M_{\odot}$ normal mass sequence}
\end{center}
\end{table}

\begin{table}[h]
\begin{center}
\begin{tabular}{|c|c|c|c|c|c|c|}
$\Omega (s^{-1})$ & $R_{num}(km)$ & $R_{calc}(km)$ & $\frac{R_{calc}}{R_{num}}$ & $%
M_{num}(M_{\odot })$ & $M_{calc}(M_{\odot })$ & $\frac{M_{calc}}{M_{num}}$ \\\hline 
0 & 9.411 & 9.411 & 1.0 & 2.1335 & 2.1335 & 1.0 \\\hline 
3784.6 & 9.744 & 9.669 & 0.992 & 2.1417 & 2.174 & 1.015 \\\hline 
5827.1 & 10.05 & 10.060 & 1.001 & 2.1547 & 2.216 & 1.028 \\\hline 
7381.3 & 10.40 & 10.525 & 1.012 & 2.1703 & 2.304 & 1.061 \\\hline 
8507.7 & 10.78 & 10.987 & 1.019 & 2.1863 & 2.371 & 1.084 \\\hline 
9308.9 & 11.20 & 11.403 & 1.018 & 2.2018 & 2.430 & 1.104 \\\hline 
9854.5 & 11.64 & 11.740 & 1.008 & 2.2157 & 2.477 & 1.118 \\\hline 
10211.0 & 12.10 & 11.988 & 0.990 & 2.2278 & 2.512 & 1.127 \\\hline 
10426.0 & 12.57 & 12.150 & 0.966 & 2.2373 & 2.534 & 1.132 \\\hline 
10587.0 & 13.66 & 12.279 & 0.898 & 2.2467 & 2.551 & 1.135
\end{tabular}
\caption{Comparison of the radius $R_{calc}$ and mass $M_{calc}$ obtained
from Eqs.(\ref{b})-(\ref{bbb}) with the
radius $R_{num}$ and mass $M_{num}$ obtained by numerically integrating the gravitational field
equations [20] for EOS AU [23]: maximum mass normal mass sequence}
\end{center}
\end{table}

\begin{table}[h]
\begin{center}
\begin{tabular}{|c|c|c|c|c|c|c|}
$\Omega (s^{-1})$ & $R_{num}(km)$ & $R_{calc}(km)$ & $\frac{R_{calc}}{R_{num}}$ & $%
M_{num}(M_{\odot })$ & $M_{calc}(M_{\odot })$ & $\frac{M_{calc}}{M_{num}}$ \\\hline 
0 & 13.70 & 13.70 & 1.0 & 2.7002 & 2.7002 & 1.0 \\\hline 
2212.2 & 14.20 & 14.01 & 0.986 & 2.7063 & 2.739 & 1.012 \\\hline 
3495.1 & 14.68 & 14.52 & 0.989 & 2.7181 & 2.800 & 1.030 \\\hline 
4485.0 & 15.24 & 15.13 & 0.993 & 2.7331 & 2.869 & 1.050 \\\hline 
5212.1 & 15.88 & 15.75 & 0.992 & 2.7494 & 2.936 & 1.068 \\\hline 
5712.7 & 16.57 & 16.29 & 0.983 & 2.7650 & 2.991 & 1.081 \\\hline 
6051.2 & 17.30 & 16.72 & 0.966 & 2.7797 & 3.033 & 1.091 \\\hline 
6263.5 & 18.05 & 17.02 & 0.942 & 2.7923 & 3.062 & 1.096 \\\hline 
6422.3 & 19.14 & 17.26 & 0.901 & 2.8056 & 3.086 & 1.099 \\\hline 
6482.9 & 20.66 & 17.36 & 0.840 & 2.8126 & 3.095 & 1.100
\end{tabular}
\caption{Comparison of the radius $R_{calc}$ and mass $M_{calc}$ obtained
from Eqs.(\ref{b})-(\ref{bbb}) with the
radius $R_{num}$ and mass $M_{num}$ obtained by numerically integrating the gravitational field
equations [20] for EOS L [24]: maximum mass normal mass sequence}
\end{center}
\end{table}

For $1.4M_{\odot}$ sequences described by EOS A and FPS the maximum
error of our prediction is around $9\%$. For the maximum mass normal sequences
 of EOS AU and L the maximum
error in the predicted value of the mass does not exceed $14\%$ but for
large angular speeds it is around $16\%$ for the radius of the rotating compact object.

\section{APPLICATIONS TO STRANGE STARS}

It is generally believed today that strange quark matter,consisting of u-,d-
and s quarks is energetically the most favorable state of quark matter.
Witten [27] suggested that there are two ways of formation of the strange
matter: the quark-hadron phase transition in the early universe and
conversion of neutron stars into strange ones at ultrahigh densities.In the
theories of strong interactions quark bag models suppose that the breaking
of physical vacuum takes place inside hadrons. As a result the vacuum energy
densities inside and outside a hadron become essentially different and the
vacuum pressure on a bag wall equilibrates the pressure of quarks thus
stabilizing the system.

If the hypothesis of the quark matter is true, then some of neutrons stars
could actually be strange stars, built entirely of strange matter [28],[29].
Caldwell and Friedman [30] have presented arguments against the existence of
strange stars. For a recent review of strange star properties see [31].

There are several proposed mechanisms for the formation of quark stars after
galaxy formation. Strange stars are expected to form during the collapse of
the core of a massive star after the supernova explosion as a result of a
first or second order phase transition, resulting in deconfined quark matter
[32].Another possibility for strange star formation is that some rapidly
spinning neutron stars in low-mass X-ray binaries (LXMBs) can accrete
sufficient mass to undergo a phase transition to become strange stars [33].
In this scenario it is supposed that at the beginning of accretion the mass
of the neutron star is 1.4. It has been shown [34] that the amounts of
matter accreted by 18 millisecond pulsars in binary systems exceed $%
0.5M_{\odot }$. Hence some of the millisecond pulsars may be strange
stars.Strange stars have also been proposed as sources of unusual
astrophysical phenomena, e.g. soft $\gamma$-ray repeaters [35],pulsating
X-ray burster [36], cosmological $\gamma$-ray bursts [35], [37-38] etc.The
mechanism of the phase transition from neutron to quark stars in low LXMBs
also results in the excitation of stellar radial oscillations that can be
damped by gravitational wave radiation instead of internal viscosity [39]
.The discovery of kHz quasi-periodic oscillation in LXMBs [40] implies that
the compact stellar object must have very soft equation of state, which is
consistent with that of strange stars [41-42].

Assuming that interactions of quarks and gluons are sufficiently small the
energy density $\rho $ and pressure $p$ of a quark-gluon plasma at
temperature $T$ and chemical potential $\mu_f$ (the subscript f denotes the
various quark flavors u,d,s etc.) can be calculated by thermal theory.
Neglecting quark masses in first order perturbation theory and supposing
that quarks are confined to the bag volume (in the case of a bare strange
star, the boundary of the bag coincides with stellar surface), the equation
of state is 
\begin{equation}\label{q}
p=\frac{(\rho-4B)}{3},
\end{equation}
where $B$ is the difference between the energy density of the perturbative
and non-perturbative QCD vacuum (the bag constant). Equation (\ref{q}) is
essentially the equation of state of a gas of massive particles with
corrections due to the QCD trace anomaly and perturbative interactions.These
are always negative , reducing the energy density at given temperature by
about a factor two [43]. In the limit $p\to 0$ (at the star's surface) we
have $\rho\to 4B$. The equation of state (\ref{q}) does not depend upon
quark flavor number,hence it will be correct either for strange quark matter
($m_s\to 0$) or for normal quark matter ($m_s\to\infty$). For any
intermediate values of $m_s$ the state equation (\ref{q}) gives the pressure
with error less than 4\% [31].Thus the equation of state of strange matter
is mainly determined by the vacuum energy density $B$.

The bag model equation of state (\ref{q}) has been the basis for the study
of most of the static relativistic models of strange stars [27],[29]. Based
on the numerical integration of the mass continuity and hydrostatic
equilibrium TOV (Tolman-Oppenheimer-Volkoff) equations for different values
of the bag constant these authors obtained a complete description of static
strange stars. Using numerical methods Witten [27] and Haensel et al. [29]
obtained the maximum gravitational mass $M_{max}$, the maximum baryon mass $%
M_{B,max}\equiv1.66\times10^{-27}kg\times N_B$ ($N_B$-the total baryon
number of the stellar configuration) and the maximum radius $R_{max}$ of the
strange star ,as a function of the bag constant, in the form [27],[29],[9]: 
\begin{equation}\label{q1}
M_{max}=\frac{1.9638M_\odot }{\sqrt{B_{60}}},M_{B,max}=\frac{2.6252M_\odot }{%
\sqrt{B_{60}}}, R_{max}=\frac{10.172km}{\sqrt{B_{60}}},
\end{equation}
where $B_{60}\equiv B/(60MeVfm^{-3})$.

Colpi and Miller [44] and Glendenning and Weber [45] have investigated the
rotational properties of strange stars in the slow rotation approximation.
As far as rotational deformations are concerned, there are a number of
detailed differences between the strange star models and standard neutron
stars. Exact numerical calculations of rapidly rotating strange stars were
done by Lattimer et al. [10], Gourgoulhon et al.[9] (by using a multi-domain
spectral method that enable to treat exactly the density discontinuity at
the surface of strange stars) and by Stergioulas,Kluzniak and Bulik [42].
Rotation increases maximum allowable mass of strange stars and the
equatorial radius of the maximum mass configuration. Gourgoulhon et al. [9]
obtained for the maximum mass and radius of quark stars the following two
exact formulae 
\begin{equation}\label{q2}
M_{max}^{rot}=\frac{2.831M_\odot }{\sqrt{B_{60}}},R_{eq,M_{max}}^{rot}=\frac{%
16.54km}{\sqrt{B_{60}}}.
\end{equation}

In the present section we shall derive, by using the formalism presented in
Section I, analytic mass and radius formulae for general relativistic static
and rotating equilibrium strange matter configurations described by the bag
model equation of state (\ref{q}).We shall begin with the study of the
static strange star, but presenting and alternative and physically more
involved discussion of this case.

The changes caused by the general theory of relativity in the conditions of
thermal equilibrium,taking into account the gravitational field of the
body,are of fundamental importance.In a constant gravitational field we must
distinguish the conserved energy $E_0$ of any small part of the stellar
object from the energy $E$ measured by an observer situated at a given
point. These two quantities are related by $E_0=E\sqrt{g_{00}}$ [46], where $%
g_{00}$ is the time component of the metric tensor. A similar change occurs
in the condition of the constancy of the chemical potential throughout the
star. The chemical potential is defined as the derivative of the energy with
respect to the number of particles $N$, $\mu=(\frac{\partial E}{\partial N}%
)_{S,V}$.Since this number is a constant for the stellar object, $N=constant$%
,for the chemical potential measured at any point inside the gravitating
body we have the relation [46]: 
\begin{equation}\label{q3}
\mu\sqrt{g_{00}}=constant.
\end{equation}

A similar relation also holds for the temperature $T$, $T\sqrt{g_{00}}%
=constant.$, since we suppose that the strange star is in thermal
equilibrium [46].Consequently, $\frac{\mu}{T}=constant$ inside the compact
object. Hence $\frac{d\mu}{\mu}=\frac{dT}{T}$.At constant volume (equal to
unity) we have $dp=sdT+nd\mu$,where $s$ and $n$ are the entropy and number
of particles in unit volume of the body,respectively. With the use of $dT=T%
\frac{d\mu}{\mu}$ and taking into account that $\mu n+sT=\rho +p$ we obtain
the following equation relating the equilibrium chemical potential to the
energy density and pressure of the star [46]: 
\begin{equation}\label{q4}
\frac{d\mu}{\mu}=\frac{dp}{\rho+p}.
\end{equation}

Consider now a static equilibrium quark matter configuration satisfying the
bag model equation of state (\ref{q}). Let us compare the values of the
chemical potential $\mu =\sum \mu _{f}$ at two points:at the center of the
star and at the vacuum boundary. From equation (\ref{q3}) we obtain 
\begin{equation}\label{q41}
\left[ \mu \sqrt{g_{00}}\right] _{center}=\left[ \mu _{S}\sqrt{g_{00}}%
\right] _{boundary},  
\end{equation}
where the indices $C$ and $S$ refer to the center and to surface of quark
star respectively.At the vacuum boundary the gravitational field of the
strange star is described by the Schwarzschild solution, which gives \lbrack
17\rbrack : 
\begin{equation}\label{q5}
g_{00}=1-\frac{2M_{stat}}{R_{stat}}, 
\end{equation}
where $M_{stat}$ and $R_{stat}$ are the total mass and radius of the static
strange star,respectively.At the center of the star the time component of
the metric tensor has a constant value (this also follows from the Bianchi
identity $\frac{d}{dr}\ln (g_{00})=-\frac{2}{\epsilon +p}\frac{dp}{dr}$
\lbrack 17\rbrack\ and we denote 
\begin{equation}\label{q6}
g_{00}\mid _{center}=C\left( \rho _{C},B\right) . 
\end{equation}

From a physical point of view $C(\rho _{C},B)$ can be related via the
relation $\frac{1}{C}-1=z_{c}$ to the redshift $z_{c}$ of a photon emitted
from the center of the quark star. For a given static strange matter
configuration the value of $C$ depends only on the central density of the
quark star $\rho _{C}$ and on the bag constant. Therefore from equation (\ref
{q41}) we obtain 
\begin{equation}\label{q7}
\frac{\mu _{C}^{2}}{\mu _{S}^{2}}=\frac{1}{C\left( \rho _{C},B\right) }%
\left( 1-\frac{2M_{stat}}{R_{stat}}\right) .  
\end{equation}
With the use of the bag model equation of state (\ref{q}) we can integrate
equation (\ref{q4}) to obtain 
\begin{equation}
\mu =C_{0}(\rho -B)^{1/4}.
\end{equation}

The integration constant $C_0$ can be determined by calculating the chemical
potential at the center of the quark star. Hence we obtain 
\begin{equation}
C_0=\frac{\mu_C}{(\rho_C-B)^{1/4}}.
\end{equation}

The variation of the chemical potential inside the quark star can be
represented as 
\begin{equation}\label{q8}
\mu=\frac{\mu_C}{(\rho_C-B)^{1/4}}(\rho-B)^{1/4}.
\end{equation}

At the surface of the star $\rho_S\to4B$.Therefore from equation (\ref{q8})
it immediately follows that 
\begin{equation}\label{q9}
\mu_S=(3B)^{1/4}\frac{\mu_C}{(\rho_C-B)^{1/4}}.
\end{equation}

In order to simplify notation we shall introduce a dimensionless parameter $%
\eta =\frac{\rho _{C}}{B}$, so that $C(\rho _{C},B)=C(\eta )$.By eliminating 
$\frac{\mu _{C}}{\mu _{S}}$ from equations (\ref{q7}) and (\ref{q9}) we
obtain the following exact formula for the mass-radius ratio of a strange
star: 
\begin{equation}\label{q10}
\frac{M_{stat}}{R_{stat}}=\left[ 1-C(\eta )\left( \frac{\eta -1}{3}\right) ^{%
\frac{1}{2}})\right] .
\end{equation}

For a given equation of state the mass-radius ratio of the star depends on
the values of the metric tensor component at the center of the star,$C(\eta)$%
, only.A possible representation for the function giving the values of $%
g_{00}$ at the center of the quark star is in the form of a power series $%
C(\eta)=const.+\sum\frac{a_i}{\eta^{i}}$,with $a_i$ constants.

As applied on the star surface the mass continuity equation leads to a rough
approximation of the quark star mass of the form $\frac{dM_{stat}}{dr}%
\approx \frac{M_{stat}}{R_{stat}}\sim 16\pi BR_{stat}^2$. A mass-radius
relation of this form could also describe zero pressure quark matter,with $%
\epsilon=4B$,$p\equiv 0$ and $M_{stat}(R_{stat})=\frac{16\pi BR_{stat}^3}{3}$%
. But for densities greater than $4B$ the effects determined by the large
central density become important. Hence for strange quark stars we propose
the following mass-radius relation: 
\begin{equation}\label{q11}
M_{stat}=\frac{16\pi B}{3}a(\eta)R_{stat}^3,
\end{equation}
with $a(\eta)$ a function describing the variation in the quark star mass
due to the increase of the central density.

The exact form and the values of the functions $C(\eta)$ and $a(\eta)$ can
be determined only by numerically integrating the gravitational field
equations.By fitting the numerical data given in [47] for the mass and
radius of the strange star with the expressions (\ref{q10}) and (\ref{q11})
we obtain the following representations for these functions (in the present
paper we shall consider $B=10^{14}g/cm^3=56Mev/fm^3$): 
\begin{equation}\label{q12}
C(\eta)=44.005\frac{1}{\eta^3}-6.68158\frac{1}{\eta^2}+2.7403\frac{1}{\eta}%
+0.0554667,
\end{equation}
\begin{equation}\label{q13}
a(\eta)=0.0000521833\eta^3-0.00378523\eta^2+0.114564\eta+0.624094.
\end{equation}

The numerical constants in equations (\ref{q12}) and (\ref{q13}) depend on $%
B $ because the numerical data have been calculated at a given $B$. For the
polynomial fittings (\ref{q12}) and (\ref{q13}) the correlation coefficient $%
r=0.9997$ and the probability $P<0.001$. Therefore for a given value of the
bag constant $B$ we obtain the following exact representations for the
radius and mass of the static strange matter configuration obeying the MIT
bag model equation of state: 
\begin{equation}\label{q14}
R_{stat}(\eta )=\left( \frac{3c^{2}}{32\pi GB}\right) ^{1/2}\left( \frac{1}{%
a(\eta )}\left( 1-C(\eta )\left( \frac{\eta -1}{3}\right) ^{\frac{1}{2}%
}\right) \right) ^{\frac{1}{2}},
\end{equation}
\begin{equation}\label{q15}
M_{stat}(\eta )=\frac{\sqrt{3}}{2}\left( \frac{c^{2}}{32\pi GB}\right) ^{%
\frac{1}{2}}\left( a(\eta )^{-\frac{1}{2}}\left( 1-C(\eta )\left( \frac{\eta
-1}{3}\right) ^{\frac{1}{2}}\right) \right) ^{\frac{3}{2}}.
\end{equation}

The variations of the radius and mass for a strange star ($B=10^{1}4g/cm^{3}$%
) as a function of the parameter $\eta $ are represented in Figures 1 and 2.
For the sake of comparison we have also presented the data obtained by
numerically integrating the TOV and hydrostatic equilibrium equations
\lbrack 47\rbrack .Using (\ref{q12})-(\ref{q15}) we can reproduce the values
of the mass and radius of the quark star obtained by numerical integration
with an error smaller than 1\%. The maximum radius of the strange star is
obtained from the condition $\frac{dR}{d\eta }=0$.The corresponding
algebraic equation has the solution $\eta _{max}^{R}=9.99012$ (this value
depends of course on the value of $B$),giving the value of the ratio of the
central pressure and bag constant for the maximum allowable radius $R_{max}$
of the static strange star.This can be expressed as 
\begin{equation}\label{q16}
R_{max}^{stat}=0.569906\times \sqrt{\frac{3c^{2}}{32\pi BG}}, 
\end{equation}
and its numerical value for $B=10^{14}g/cm^{3}$ is $R_{max}=1.1436\times
10^{6}cm$.From the condition $\frac{dM}{d\eta }=0$ it follows that $\eta
_{max}^{(M)}=22.41173$ and the maximum mass of the static quark star is
given by 
\begin{equation}\label{q17}
M_{max}^{stat}=0.297866\times \frac{\sqrt{3}}{2(32\pi B)^{\frac{1}{2}}}%
\left( \frac{c^{2}}{G}\right) ^{\frac{3}{2}}.  
\end{equation}

From equation (\ref{q17}) and for the chosen value of the bag constant,we
obtain a value of $M_{max}=2.016M_{\odot}$.These results are in good
agreement with the previously proposed (Witten [27],Haensel,Zdunik and
Schaeffer [29]) maximum radius and mass values,given by equations (\ref{q1})
(from equations (\ref{q1}) and for $B=56MeV/fm^3$ we obtain $%
M_{max}=2.03M_{\odot}$).For values of $\eta>\eta_{max}^{(M)}$ static quark
star models would be unstable to radial perturbations.

As an application of the mass and radius formulae obtained for the static
strange stars we shall derive an explicit expression for the total energy of
the quark star. The total energy (including the gravitational field
contribution) inside an equipotential surface $S$ can be defined, according
to Lynden-Bell and Katz \lbrack 48\rbrack\ and Gron and Johannesen \lbrack
49\rbrack\ to be 
\begin{equation}\label{q18}
E=E_{M}+E_{F}=\frac{1}{8\pi }\xi _{s}\int_{S}\lbrack K\rbrack ds,
\end{equation}
where $\xi ^{i}$ is a Killing vector field of time translation, $\xi _{s}$
its value at $S$ and $\lbrack K\rbrack $ is the jump across the shell of the
trace of the extrinsic curvature of $S$, considered as embedded in 2-space $%
t=constant$.$E_{M}=\int_{S}T_{i}^{k}\xi ^{i}\sqrt{-g}dS_{k}$ and $E_{F}$ are
the energy of the matter and of the gravitational field,respectively. This
definition is manifestly coordinate invariant. In the case of the static
strange star with the use of equation (\ref{q14})and (\ref{q15}) we obtain
for the total energy (also including the gravitational contribution) the
following exact expression: 
\begin{equation}\label{q19}
E=E_{SQM}+E_{F}=-\sqrt{\frac{3}{32\pi B}}C(\eta )\left[ \frac{1}{a(\eta )}%
\frac{\eta -1}{3}\left( 1-C(\eta )\sqrt{\frac{\eta -1}{3}}\right) \right]
^{1/2},
\end{equation}
where $E_{SQM}$ is the total energy of the quark matter.

The variation of the total energy of the strange star as a function of the
parameter $\eta$ is represented in Figure 3.

The minimum value of the total-matter plus gravitational-energy of the
strange matter configuration is obtained for $\eta_{min}=5.68171$.The most
stable static stellar configuration made of strange matter is given by quark
stars with radius $R_{stab}=9.97179\times10^5cm$ and with mass $%
M_{stab}=0.96558M_\odot$, corresponding to values of the central density of
the order of $\rho_C=5.681\times B=5.681\times 10^{14}g/cm^3$.

We shall consider now the study of the rotating strange star
configurations. We shall compare our results obtained with the use of equations
(\ref{b})-(\ref{bbb}) and (\ref{q14})-(\ref{q15}) with the results provided by
Stergioulas,Kluzniak and Bulik \lbrack 42\rbrack\ and obtained by
numerically integrating the gravitational field equations for maximally
rotating (''Keplerien'') models of strange stars.The results of Stergioulas
et al.\lbrack 42\rbrack\ are also in very good agreement with the results of
the exact numerical models of rotating strange stars built of self bounded
quark matter of Gourgoulhon et al. \lbrack 9\rbrack , the difference between
these two works being smaller than $1\%$. In order to improve the accuracy
of the expressions (\ref{b})-(\ref{bbb}) we shall consider that the function $C_{rot}^{(2)}\left( \eta ,R_{e},\Omega \right)$
can be expressed in a more general form as
\begin{equation}
C_{rot}^{(2)}\left( \eta ,R_{e},\Omega \right) =\left( 1-R_{e}^{2}\Omega
^{2}\right) ^{\gamma }C\left( \eta \right),
\end{equation}

and we will assume that the parameters $\alpha $ and $\gamma $
are not constants, but some angular velocity dependent functions given by
\begin{equation}\label{q23}
\alpha =\frac{1.23188\times 10^{9}}{\Omega ^{2}}-\frac{2.15445\times 10^{5}%
}{\Omega }+13.1399,
\end{equation}
\begin{equation}\label{q24}
\gamma =9.66251\times 10^{11}\Omega ^{-3}-4.56952\times 10^{8}\Omega
^{-2}+6.99642\times 10^{4}\Omega ^{-1}-2.54401.
\end{equation}

Equations (\ref{q23})-(\ref{q24}) take into account the variation of the central density of the
maximally rotating strange star due to the increase of the angular velocity.

In Fig.4 we have represented the variation of the radius of the strange
stars,given by equation (\ref{q22}) together with the angular velocity dependent functions $\alpha $ and 
$\gamma $ and the values given in Stergioulas et al.\lbrack 42\rbrack ,
calculated for the same values of the central density and angular
velocity.The mean of the difference between these two sets of values is
smaller than $1\%$. The variation of the radius of the maximally rotating strange star as
a function of both central density and angular velocity is represented in
Fig.5.Figs.6 and 7 present the variation of the mass of the rapidly rotating
strange star as a function of $\ $\ the angular velocity $\Omega $ and of $\
\Omega $ and central density, respectively.

From equation (\ref{b}) and equations (\ref{q14}) and (\ref{q15})
it follows that the radius of the rotating strange star can be
expressed, as a function of the central density and angular velocity only,in
the following form:
\begin{equation}\label{q22}
R_{e}=\sqrt{\frac{12\left( 1-C(\eta )\sqrt{\frac{\eta -1}{3}}\right) }{64\pi
Ba(\eta )-3\alpha \Omega ^{2}}}. 
\end{equation}

Hence in this approximation the basic rotational parameters of maximally
rotating strange star can
be represented in terms of the static configuration and of the angular
velocity only.

\section{DISCUSSIONS AND FINAL REMARKS}

In the present paper we have suggested the possibility of the existence of
a universal pattern expressing the basic properties of rotating 
compact object as simple functions of the parameters of the
static object and of the angular velocity only.
We have obtained exact formulae which give the dependence
of the radius and mass of the static and rotating stars
on the central density of the stellar object
and of its angular velocity.In the static case this is made possible
due to the constancy of the chemical potential.
The two unknown functions involved
in the model must be obtained by fitting the exact formulae with data obtained
from the numerical integration of the structure equations of the neutron or quark star.
The resulting analytical expressions can reproduce the radius and mass of the strange star with an
error smaller than $1\%$ and they also provide a simple way to obtain the maximum
mass and radius of the static configuration.

In the rotating case,with the use of the hydrostatic equilibrium condition,which is
the consequence of the Bianchi identities, we have also obtained exact mass-radius
relations,depending on three functions describing the effect of rotation on star
structure.These relations are exact in the sense that they have been obtained without
any special assumptions.By assuming some appropriate forms for the unknown functions
we have obtained a general description of the mass and radius of the rotating neutron or strange stars,
which generally and for a broad class of equations of state can reproduce the values
obtained by numerical integration of the gravitational
field equations with a mean error of around $5\%$.The expressions of the unknown parameters
have been chosen following a close analogy with the static case,whose relevance for
the study of rotating general relativistic configurations seems to be more important
than previously believed.These functions also incorporate some other general relativistic
effects not explicitly taken into account,like the variation of the moment of inertia
of the star with the angular velocity. Ravenhall and Pethick [19] have presented
a formula valid for a broad range of realistic equations of state of dense matter
expressing the moment of inertia in terms of stellar mass and radius. We have not used
these results,obtained in the slow rotation limit because,at least in the case of strange stars,
the Newtonian expression of the moment of inertia also leads to a quite accurate
physical description of the rotating objects.As an application of the obtained
formulae we had given a derivation of the "empirical" formula relating
the maximum angular velocity to the mass and radius of the static maximal stellar configurations.

The possibility of obtaining the basic parameters of general relativistic rotating
objects in terms of static parameters could lead to a major computational simplification
in the study of rotation.The relation between the presented formalism and the
Einstein gravitational field equations will be the subject of a future publication.

\begin{figure}
\vbox to7.2in{\rule{0pt}{7.2in}}
\includegraphics{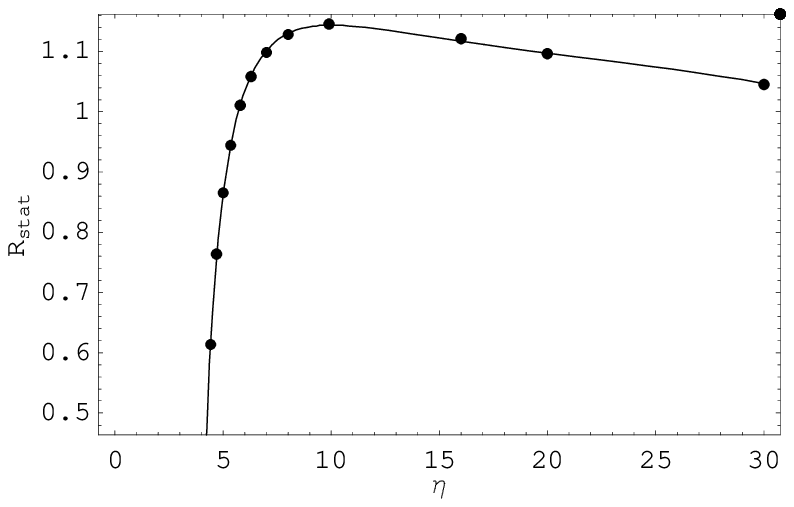}
\caption{Variation of the radius (in units of $10^6cm$) of the static strange star as a function of
the parameter $\eta$. The solid curve represents the values of the
radius calculated with equation (\ref{q14}),while the points represent
the values of the radius of the quark star obtained by numerically
integrating the general relativistic structure equations of the star.}
\label{FIG1}
\end{figure}

\begin{figure}
\vbox to7.2in{\rule{0pt}{7.2in}}
\includegraphics{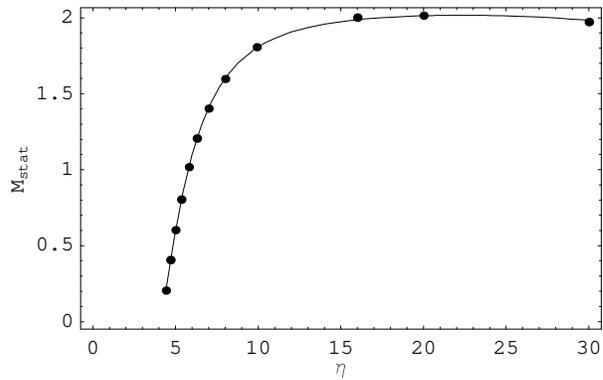}
\caption{Variation of the mass of the static strange star(in solar mass units) as a function of the
parameter $\eta$.The solid curve represents the values of the mass
calculated with equation (\ref{q15}), while the points represent the values
of the mass of the strange star obtained by numerically integrating
the general relativistic structure equations of the star.}
\label{FIG2}
\end{figure}

\begin{figure}
\vbox to7.2in{\rule{0pt}{7.2in}}
\includegraphics{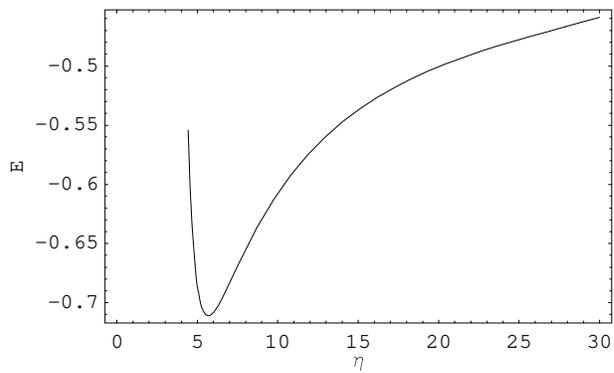}
\caption{Variation of the total energy (matter+gravitational) of the
static strange quark star.}
\label{FIG3}
\end{figure}

\begin{figure}
\vbox to7.2in{\rule{0pt}{7.2in}}
\includegraphics{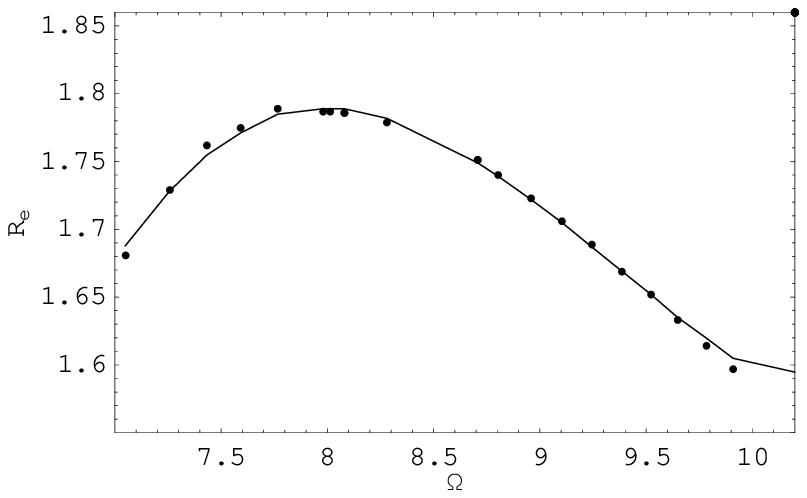}
\caption{The radius of rotating strange star (in units of $10^6cm$)
,as a function of the angular velocity, for values of the central
density in the range $\eta\in[5,16.63]$.Values obtained from the
analytical expression (solid curve) are compared with data obtained
by Stergioulas et al. [42] numerically integrating the gravitational
field equations (points).}
\label{FIG4}
\end{figure}

\begin{figure}
\vbox to7.2in{\rule{0pt}{7.2in}}
\includegraphics{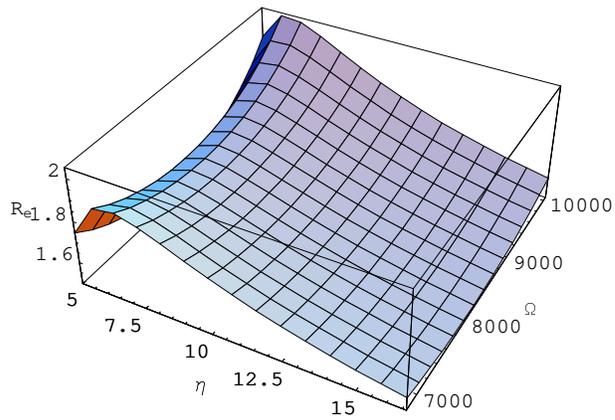}
\caption{Variation of the radius of the rotating strange star
(in units of $10^6cm$) as a function of the angular velocity $\Omega (s^{-1}$ 
and of the central density $\eta $.}
\label{FIG5}
\end{figure}

\begin{figure}
\vbox to7.2in{\rule{0pt}{7.2in}}
\includegraphics{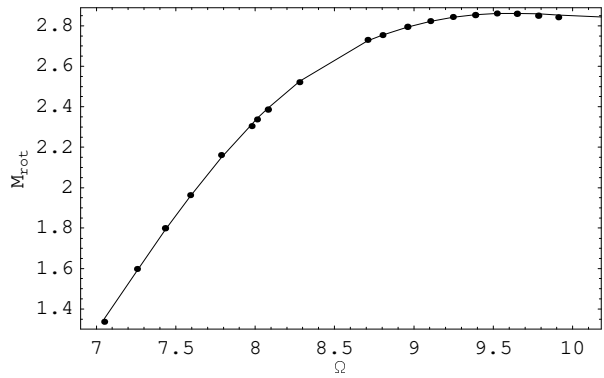}
\caption{The mass of rotating strange star,expressed in solar mass
units,as a function of the angular velocity, for values of the central
density in the range $\eta\in[5,16.63]$.Values obtained from the
analytical expression (solid curve) are compared with data obtained
by Stergioulas et al. [42] numerically integrating the gravitational
field equations (points).}
\label{FIG6}
\end{figure}

\begin{figure}
\vbox to7.2in{\rule{0pt}{7.2in}}
\includegraphics{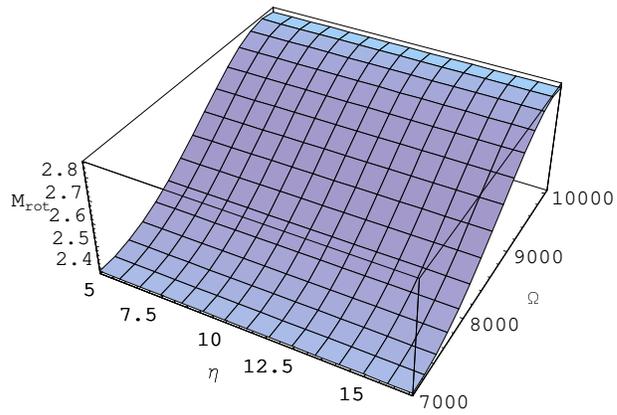}
\caption{Variation of the mass of the rotating strange star,expressed in solar mass
units,as a function of the angular velocity $\Omega $ and of the central
density $\eta $.}
\label{FIG7}
\end{figure}

\end{document}